\documentclass[runningheads]{llncs}
\usepackage[T1]{fontenc}
\usepackage{graphicx}
\usepackage{xcolor}
\usepackage{wrapfig}
\usepackage{subfigure}
\usepackage{enumitem}
\usepackage{multirow}
\usepackage{tcolorbox}
\usepackage{footmisc}
\usepackage{lineno}
\usepackage{tikz}
\newcommand*\circled[1]{\tikz[baseline=(char.base)]{\node[shape=circle,draw,inner sep=1pt] (char) {#1};}}
\def\checkmark{\tikz\fill[scale=0.4](0,.35) -- (.25,0) -- (1,.7) -- (.25,.15) -- cycle;} 
\definecolor{anti-flashwhite}{rgb}{0.95, 0.95, 0.96}
\definecolor{antiquewhite}{rgb}{0.98, 0.92, 0.84}
\begin{document}
\title{CommonUppRoad: A Framework of Formal Modelling, Verifying, Learning, and Visualisation of Autonomous Vehicles}
\titlerunning{Modelling, Verifying, Learning, and Visualisation of Autonomous Vehicles}
\author{Rong Gu\inst{1} \and
Kaige Tan\inst{2} \and
Andreas Holck Høeg-Petersen\inst{3} \and 
Lei Feng\inst{2} \and 
Kim Guldstrand Larsen\inst{3}}
\authorrunning{Rong Gu et al.}
%
\institute{Mälardalen University, Sweden\\
\email{rong.gu@mdu.se} \\\and 
KTH, Sweden \\\email{\{kaiget, lfeng\}@kth.se} \\\and 
Aalborg University, Denmark \\\email{\{ahhp, kgl\}@cs.aau.dk}}
\maketitle
\setcounter{footnote}{0}
\begin{abstract}
Combining machine learning and formal methods (FMs) provides a possible solution to overcome the safety issue of autonomous driving (AD) vehicles. However, there are gaps to be bridged before this combination becomes practically applicable and useful. In an attempt to facilitate researchers in both FMs and AD areas, this paper proposes a framework that combines two well-known tools, namely CommonRoad and UPPAAL. On the one hand, CommonRoad can be enhanced by the rigorous semantics of models in UPPAAL, which enables a systematic and comprehensive understanding of the AD system's behaviour and thus strengthens the safety of the system. On the other hand, controllers synthesised by UPPAAL can be visualised by CommonRoad in real-world road networks, which facilitates AD vehicle designers greatly adopting formal models in system design. In this framework, we provide automatic model conversions between CommonRoad and UPPAAL. Therefore, users only need to program in Python and the framework takes care of the formal models, learning, and verification in the backend. We perform experiments to demonstrate the applicability of our framework in various AD scenarios, discuss the advantages of solving motion planning in our framework, and show the scalability limit and possible solutions.
\keywords{Autonomous vehicles  \and Motion planning \and UPPAAL \and CommonRoad \and Reinforcement learning.}
\end{abstract}
\section{Introduction}\label{sec:intro}
\textit{Autonomous Driving (AD)} has seen a significant development in the automotive industry in the past decade.
The Society of Automotive Engineers (SAE) has defined the six levels of AD \cite{sae6levels}. Levels 0 - 2 are what we have in most of the modern vehicles nowadays.
Level-3 AD means vehicles can detect the environment and make informed decisions by themselves, but they still need human drivers to stay alert and take control at any moment, whereas Level-4 AD vehicles do not require human interaction in \textit{most} circumstances, however, humans still have the option to manually override in case of emergencies.
The highest AD level is five, where vehicles do not require human attention at all.
The current technologies can only achieve level-3 AD or level-4 AD in conceptual vehicular models.
One of the biggest challenges is the \textit{safety issue}.
Different from the casualties caused by human-driven cars, the public can hardly accept even one accident caused by an AD vehicle \cite{uberAcc}\cite{teslaAcc}.
Automotive companies are running road tests for their AD vehicles over millions of miles a year, and yet accidents still keep occurring~\cite{zhang2022finding}.
Therefore, to increase the public's confidence in AD, road-based or even simulation-based testing of AD vehicles is not enough.
Formal methods (FMs) are well-known for providing rigorous analysis of safety-critical systems. In recent years, a great amount of research has been carried out on the application of FMs in overcoming all sorts of safety issues of AD \cite{gu2022correctness}\cite{sanchez2022foresee}\cite{yang2023reinforcement}.
However, the usability and scalability of FMs when being adopted in AD systems are still challenging.
First, as mathematics-based methods, FMs need a steep learning curve for AD researchers and practitioners.
Moreover, the lack of tools like Matlab in the FMs community, which provide off-the-shelf blocks for users to reuse, means of visualization, and conversion between models and executable code, is a significant drawback. 
These disadvantages hinder users outside the FMs community from applying formal methods in their domains.
Second, when considering complex scenarios of AD, such as intersections with a large number of human-driven vehicles and complex traffic rules, FMs often fall short of scalability \cite{gu2020tamaa}.
Third, AD vehicles often involve machine-learning-based components, such as controllers that are trained by reinforcement learning (RL) \cite{sutton1999reinforcement}.
These components often require a huge amount of data for training and validation, and their control logic is different from traditional software, which makes the verification of such components extremely difficult.
In this paper, we propose a combination of tools from both communities of FMs and AD, namely UPPAAL \cite{larsen1997uppaal} and CommonRoad \cite{althoff2017commonroad}, which would improve the usability of FMs in AD development and enhance the safety aspect of AD.
CommonRoad\footnote{commonroad.in.tum.de} is an open-source toolset for AD development, testing, and visualization. 
CommonRoad contains scenarios of AD, vehicle dynamics models, and vehicle parameters.
%
%
%
Scenarios in CommonRoad can be designed manually in their GUI or converted from existing databases (e.g., highD \cite{highDdataset}). It also provides almost 4,000 scenarios converted from real-world driving scenes.
CommonRoad is written in Python and it supports various motion planners based on search and learning. 
Although CommonRoad has a drivability checker, 
it has no support from FMs and thus does not provide safety guarantees on the motion-planning results.
%
%
%

%
UPPAAL, as a state-of-the-art model checker, has added functions for RL in its recent release\footnote{UPPAAL 5 on uppaal.org}.
It supports modelling timed games, in which the behaviours of two players are formally described by timed (or hybrid) automata, which is a formalism for modelling real-time systems.
%
%
%
%
Therefore, we can model the real-time behaviours of AD and other vehicles, even describe their dynamics by ordinary differential equations (ODE), in timed games, and perform RL for motion planning in UPPAAL. 
Moreover, although this approach uses ODE and hybrid automata, it still keeps the possibility of exhaustive verification, which returns absolute true/false answers of safety.
%
%
%
However, UPPAAL does not have motion-plan visualisation, and the construction of timed games is very difficult for those who are not familiar with the formalism and tool.
In this paper, we propose a framework that combines CommonRoad and UPPAAL, namely \textit{CommonUppRoad}, in which we provide an automatic model conversion between UPPAAL and CommonRoad.
Specifically, we convert the network of roads, static obstacles, and planning goals in CommonRoad into constant variables of timed games in UPPAAL.
As constant variables, they do not occupy the state space of the formal model, which benefits the scalability greatly.
Further, we develop functions for collision and off-road detection in timed games.
These functions consider the shapes of vehicles, such as their widths, lengths, and orientations.
We also define the templates of timed games, which model the vehicle dynamics and the atomic actions for controlling the AD vehicles.
With the conversion of scenarios and our predefined timed games, one can start synthesising an AD controller in UPPAAL with little extra effort.
One of the benefits of using UPPAAL for controller synthesis is that it supports search-based \cite{behrmann2006tiga} and learning-based methods \cite{david2015uppaal}. 
Using the search-based method, one can synthesise a so-called \textit{permissive} controller, which is guaranteed to be safe, i.e., collision-free and always on the road.
Next, one can optimise the permissive controller by RL, e.g., a fast controller reaching the goal within a time frame.
Safety is ensured in the learning results as the state-action space has been restricted by the permissive controller, which excludes unsafe state-action pairs.
The safe and permissive controller is similar to the concept of a safety shield to RL \cite{alshiekh2018safe}, but it does not need the manual exclusion of unsafe actions.
%
%
%
We provide methods to transform UPPAAL strategies into 1) decision trees \cite{schilling2024safety}, which can be directly used in the Python code of CommonRoad, or 2) a set of trajectories injected back to the scenario files in CommonRoad.
This model conversion allows us to visualise UPPAAL strategies in real-world scenarios, thus facilitating the usability of learning and formal verification in UPPAAL.
In a nutshell, the contributions of our paper are as follows.
\begin{itemize}
    \item An automatic model conversion between CommonRoad and UPPAAL, including CommonRoad scenarios to UPPAAL models, and UPPAAL strategies to controllers or trajectories in CommonRoad.
    \item Vehicle dynamics described by timed-game templates associated with ODEs. The templates allow both search-based and learning-based controller synthesis, as well as exhaustive and statistical model checking.
    \item An experimental analysis of the usability and scalability of motion planning in UPPAAL. The planning is based on UPPAAL models converted from CommonRoad and uses both search-based and learning-based methods. 
\end{itemize}
The remainder of the paper is organised as follows. Section \ref{sec:preli} introduces the background knowledge for understanding this paper. Section \ref{sec:problem} describes the problem and an example of AD motion planning. Section \ref{sec:model} is about model conversion between CommonRoad and UPPAAL before Section \ref{sec:plan} introduces the two motion planning methods in UPPAAL. Section \ref{sec:exp} describes the experiments, results, and a discussion about the strengths and weaknesses of the current methods. Section \ref{sec:related} presents the related work in both AD and FMs communities. Section \ref{sec:conclusion} concludes the paper and introduces future work.
\section{Preliminaries}\label{sec:preli}
In this section, we introduce the preliminaries for understanding this paper as well as the mathematical notions. We first describe the formalism of verification, learning in UPPAAL, and the definition of decision trees that we use for representing strategies in UPPAAL. Next, we briefly introduce concepts of CommonRoad, which are used in this paper.
\subsection{UPPAAL, Timed Games, and Decision Trees}

\noindent
UPPAAL is a state-of-the-art model-checking tool for real-time systems \cite{larsen1997uppaal}. A UPPAAL model is defined as a network of Timed Automata (TA). A TA consists of finite sets of locations, edges and real-valued non-negative variables called \textit{clocks} which progress at the rate of one. Locations can be labelled with invariants over clocks, which must be true for the process to stay in that location. A location can also be marked as either `Urgent' or `Committed', which in both cases means that no time is allowed to pass while the process is in this location and with the additional requirement for the latter that the next transition must follow one of the outgoing edges from the committed location.

Transitions between locations can be guarded by clocks so as to ensure certain conditions are met before a transition can happen. When a transition occurs an update operation can affect the system state, for example by resetting some of the clocks. Further, we can define broadcast channels on edges such that when the model transitions via such an edge, a signal is sent to a set of listening edges which will trigger transitions via these edges simultaneously.

Transitions between locations can be interpreted as actions, and in a \textit{timed game} (TG) \cite{behrmann2006tiga}, we distinguish between controllable and uncontrollable actions, which in UPPAAL are represented by solid and dashed edges respectively. The first actions are controlled by an agent playing intentionally in order to accomplish some objectives. Uncontrollable actions either happen stochastically or are invoked by a purposefully antagonistic environment (the `other' player).

In a control setting, the agent must learn a winning strategy that can deal with a possibly antagonistic environment. A strategy can either involve adhering to specified safety requirements (for example, `never enter location $x$ if clock $c$ is less than 0.9') or optimizing for some objective function (for example, `get to location $y$ as fast as possible'). UPPAAL uses symbolic techniques to solve the first type of game and derives a strategy that provides a set of `safe' actions for a given state of the system. For the second type of game, UPPAAL has a reinforcement learning (RL) engine \cite{david2015uppaal} that via an online refinement scheme partitions the state space according to the expected cost of performing each action, mimicking classical Q-learning techniques. Furthermore, UPPAAL can learn the strategy for (near) optimal control while being subjugated to the safety strategy, thereby ensuring that only safe actions are considered.

As an example of a TG, consider the case where an agent has to move a certain distance $d$ within some time limit $\tau$ while spending as little energy as possible. The agent has access to two actions, move fast ($a_{\mathit{fast}}$) or move slow ($a_{\mathit{slow}}$). Moving fast is efficient in terms of covering distance, but more expensive in terms of energy. On the other hand, the precise distance travelled by each action is not fully known to the agent, as the environment can affect this in some way. For example, $a_{\mathit{fast}}$ might move the agent a distance somewhere between $0.4$ and $0.6$ and likewise for $a_{\mathit{slow}}$. The agent must devise a strategy that generally moves slowly (to preserve energy) but also avoids ending up in situations where it cannot be certain to reach $d$ within $\tau$ even if only choosing $a_{\mathit{fast}}$ from then on.

The UPPAAL tool allows us to use symbolic model checking to synthesize a strategy that both satisfies the safety requirement (i.e., move a distance of $d$ before $\tau$ time has passed) and maximises the optimization objective (i.e., spend as little energy as possible). 

The result is a decision-tree-like structure for each action, where every branch node is an axis-aligned split in one of the state dimensions. Each leaf node corresponds to a region of the state space and contains the estimated Q-value of performing that given action in this specific region. Given a concrete state $s$, the optimal action can then be chosen by consulting every action-specific decision tree and choosing the action whose tree yielded the largest Q-value for $s$.

\subsection{CommonRoad}
CommonRoad is an open-source framework designed to facilitate the development, evaluation, and benchmarking of motion planning algorithms for autonomous vehicles \cite{althoff2017commonroad}. It offers a comprehensive suite of tools and datasets, including realistic traffic scenarios, dynamic and static obstacles, and a wide range of environmental conditions. The platform supports modularity, enabling the integration of vehicle models, optimization techniques, and sensor data processing methods. CommonRoad is particularly valuable for simulating complex driving environments and evaluating the safety and efficiency of different motion planning strategies. It enables rigorous testing under diverse conditions, ensuring autonomous driving systems can effectively tackle real-world challenges.
\section{Problem Description}\label{sec:problem}
In this section, we define the problem and introduce a running example that we use for illustration throughout the paper.
\subsection{Research Questions}
Our motivation for this paper is to bridge the gap between the research areas of formal methods (FMs) and autonomous driving (AD).
Safety is one of the dominating factors that AD vehicles concern. However, AD techniques, such as machine learning, often fall short of safety guarantees and thus are not recommended by most industrial standards. 
FMs are well-known for providing safety guarantees for safety-critical systems. Still, the fact that using FMs requires a deep understanding of the underlying mathematical notions and methods makes the usability of FMs questionable.
In this paper, we propose a combination of two well-known tools in both areas, namely UPPAAL and CommonRoad, which we believe would greatly benefit the development of AD vehicles and the evolution of FMs.
First, the abundant scenarios in CommonRoad provide a set of benchmarks for evaluating the usability and scalability of FMs in the AD area, especially when integrating RL and model checking in the design of AD vehicles.
Second, the visualization of scenarios and vehicle trajectories in CommonRoad can facilitate the development of FMs-based motion-planning algorithms.
Last, FMs in UPPAAL can strengthen CommonRoad in the safety aspect of AD vehicle development and improve AD designers' understanding of the system because of the rigorous syntax and semantics that formal models provide.
To know if the combination of UPPAAL and CommonRoad can indeed achieve the expected scientific improvement, we design three research questions and answer them in the remainder of this paper.
\begin{itemize}
    \item \textbf{RQ1}: How can scenarios in CommonRoad be converted to UPPAAL models?
    \item \textbf{RQ2}: How can strategies in UPPAAL be represented as controllers in CommonRoad to control AD vehicles in the corresponding scenarios?
    \item \textbf{RQ3}: What is the limit of UPPAAL in solving the planning problems in CommonRoad?
\end{itemize}
The answer to RQ1 would tell us if it is possible to automatically transform a scenario in CommonRoad, including a network of roads, static and dynamic obstacles, and a planning problem, into a UPPAAL model that is syntactically and semantically correct, and how to achieve it.
The answer to RQ2 would give us a solution to visualise strategies in UPPAAL in a dynamic scenario. These two answers would greatly motivate AD designers to adopt FMs in system design and verification.
Since we want to utilise the search-based method in UPPAAL to synthesise absolutely safe AD controllers, scalability would be an issue.
Thanks to our previous study that integrates RL and model checking in UPPAAL \cite{gu2022correctness}, the scalability can be improved to an extent. 
However, the limit of UPPAAL when solving the planning problems in CommonRoad is still unknown, so we aim to answer it in RQ3.
\subsection{Example}
\begin{figure}[t]
\vspace{-0.0cm}
\setlength{\abovecaptionskip}{-0.0cm}   
\setlength{\belowcaptionskip}{-0.5cm} 
\centering
\subfigure[A scenario in CommonRoad]   
{
  \label{fig:commonroad-scene}%
  \includegraphics[width=0.46\columnwidth]{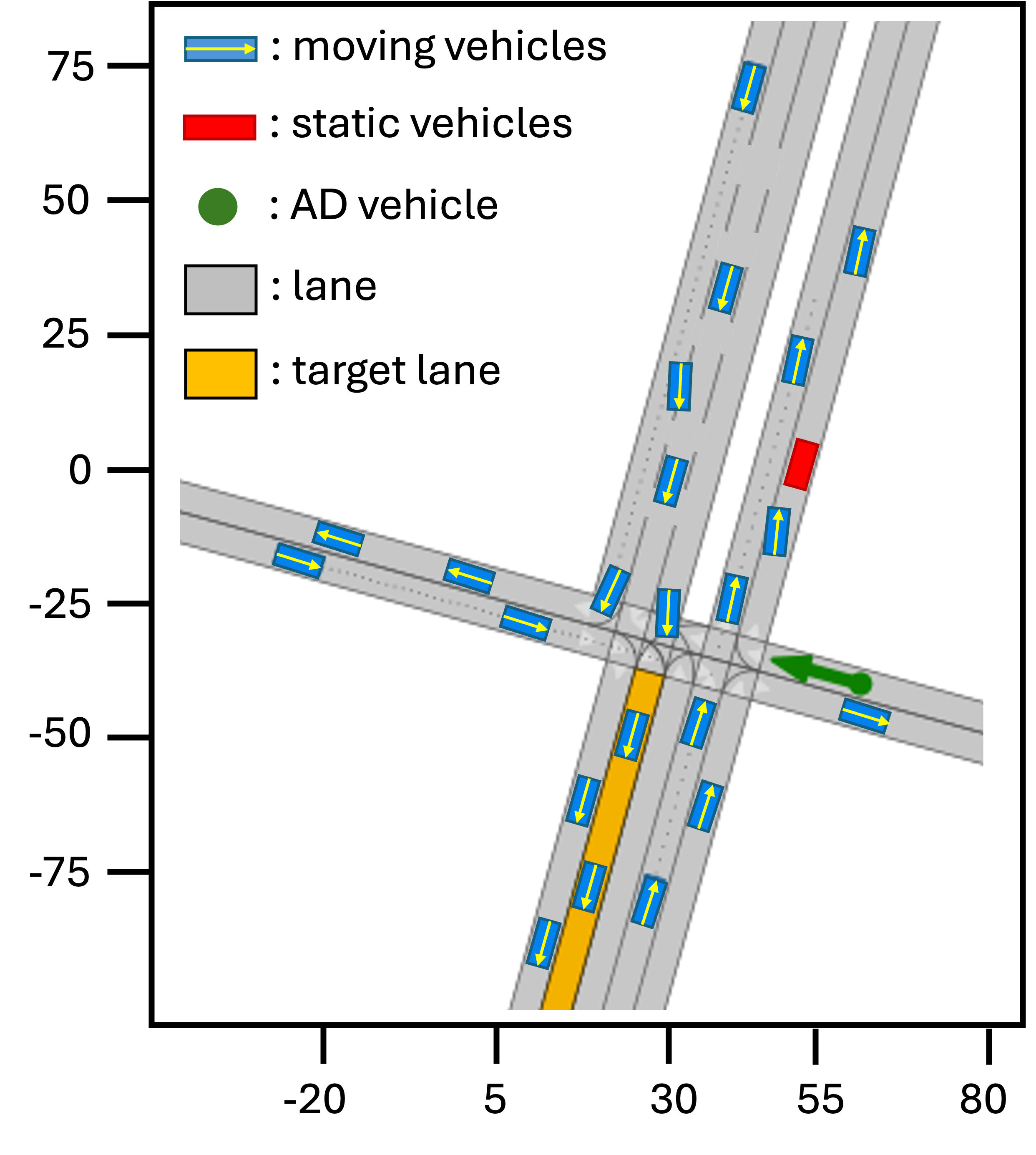}%
}
\subfigure[Structure of the scenario]{
  \label{fig:commonroad-structure}%
  \includegraphics[width=0.38\columnwidth]{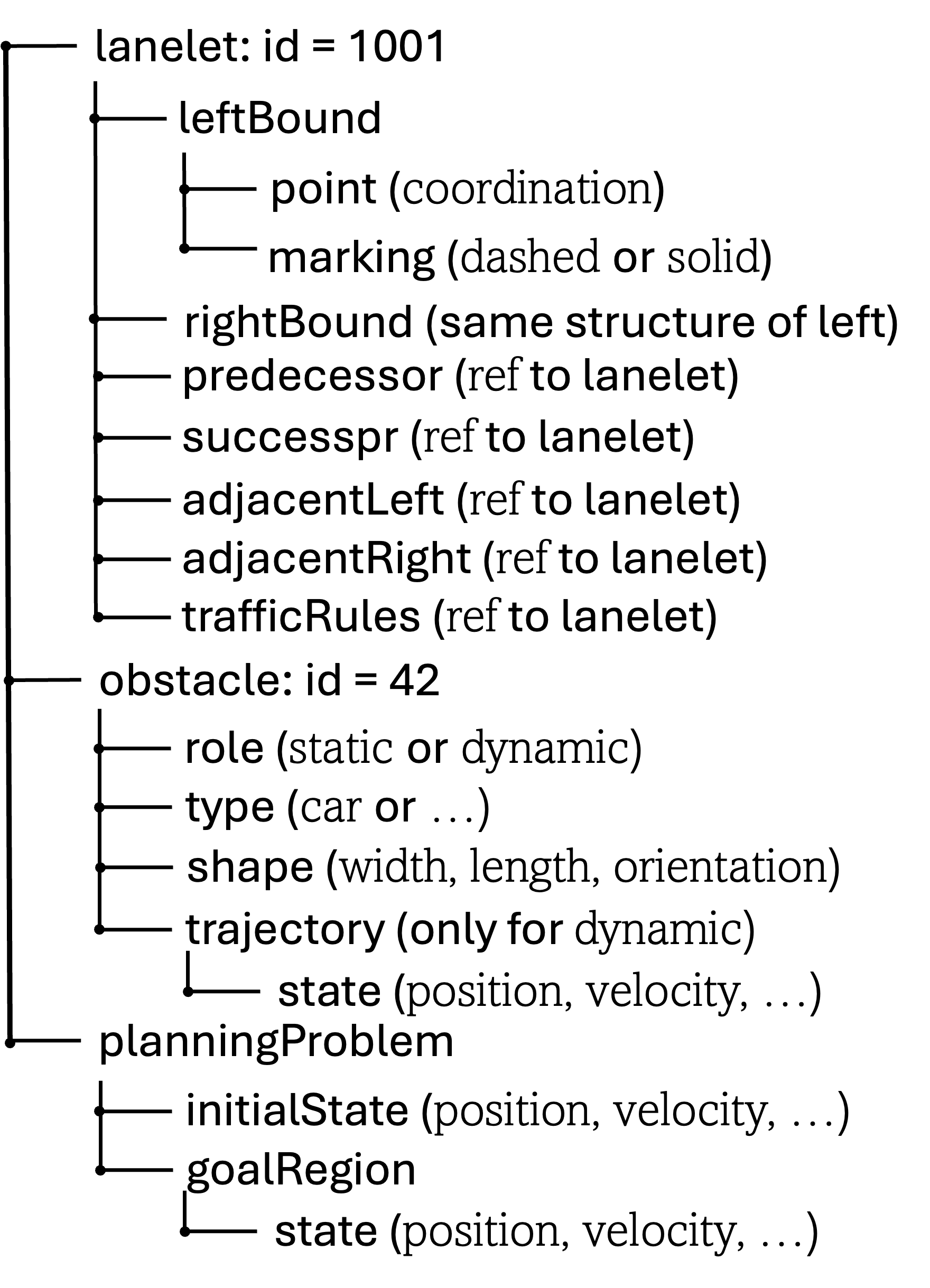}%
}
\caption{Running example.}\label{fig:commonroad-example}
\end{figure}
In this section, we introduce an example of scenarios in CommonRoad (see Fig. \ref{fig:commonroad-example}).
This example serves to illustrate the transformation of entities between UPPAAL and CommonRoad in Section \ref{sec:model}. 
Fig. \ref{fig:commonroad-scene} depicts the scenario, where an AD vehicle is travelling at an intersection and targets to turn left safely.
In this scenario, multiple lanes are separated by solid or dashed lines, representing different traffic restrictions, and other vehicles can be reactive or move along a predefined trajectory.
Fig. \ref{fig:commonroad-structure} shows the structure of the scenario XML file. As seen, the structure contains information for motion planning of the AD vehicle. CommonRoad provides methods to parse the XML file and edit the scenario in Python.
Our goal of model conversion is two-fold.
\begin{enumerate}[label=(\roman*)]
    \item Convert a scenario to a UPPAAL model for motion planning.
    \item Convert a strategy synthesised by UPPAAL into either a decision tree in Python or a moving entity with a predefined trajectory in the scenario. 
\end{enumerate}
\section{Model Conversion}\label{sec:model}
In this section, we describe our methods for model conversion between CommonRoad and UPPAAL. First, we generally describe the model conversion and the UPPAAL templates instantiated with the information converted from CommonRoad. Next, we introduce the methods of converting scenarios in CommonRoad to UPPAAL models and strategies in UPPAAL to entities in CommonRoad.
\begin{figure}[t]
    \vspace{-0.0cm}
    \setlength{\abovecaptionskip}{-0.0cm}   
    \setlength{\belowcaptionskip}{-0.5cm} 
    \centering
    \includegraphics[width=0.99\textwidth]{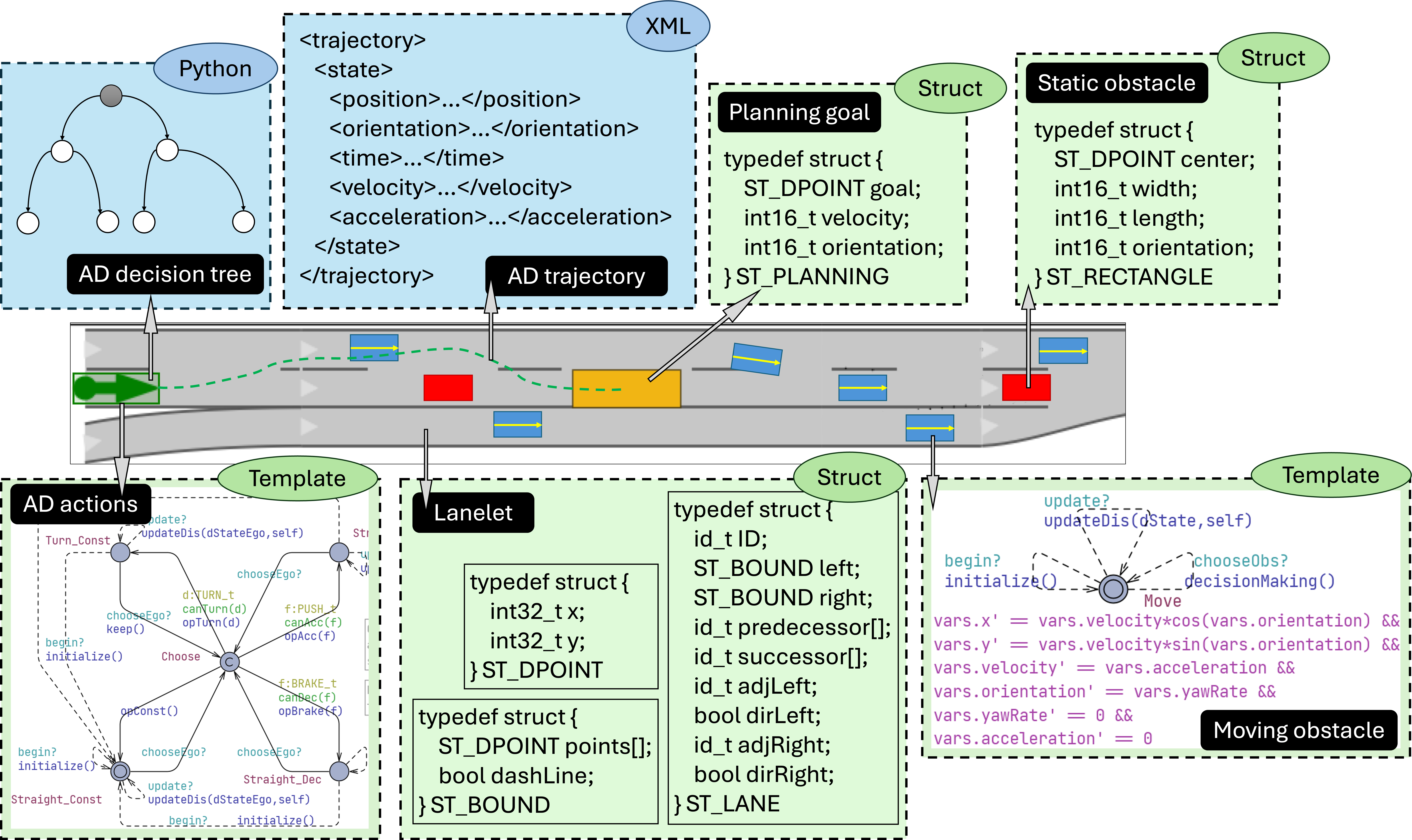}
    \caption{Overall description of model conversion. Blue boxes are entities in CommonRoad. Green boxes are entities in UPPAAL.}
    \label{fig:model_conversion}
\end{figure}
\subsection{General Description}
Fig. \ref{fig:model_conversion} generally depicts model conversion. The conversion from CommonRoad to UPPAAL can be split into two categories: i) behavioural models of the AD vehicles and moving obstacles are converted to instances of TG templates, and ii) static models of the road network (a.k.a. \texttt{lanelet}), static obstacles, and the planning goal are converted to struct variables in the C-like code of UPPAAL.
After running motion-planning in UPPAAL, we get strategies for winning the timed games. These strategies can be converted to two models in CommonRoad: i) a decision-tree controller in Python, and ii) a trajectory in the scenario XML file.
The latter can be directly visualised in CommonRoad and checked for collision and offroad by the drivability checker of CommonRoad \cite{pek2020commonroad}, whereas the former can be used as a reactive controller telling the AD vehicle what to do in different situations.
Next, we introduce the model conversion in detail.
\subsection{CommonRoad to UPPAAL}\label{sect:cr2upp}
\subsubsection{Static models: road network, obstacles, and goal.}
In CommonRoad, the scenarios are outlined in an XML file which includes a formal depiction of the road network, both static and dynamic obstacles, and the planning problem of the AD vehicle(s). Fig.~\ref{fig:commonroad-structure} provides an overview of the XML files' structure to describe a scenario. To be specific, the road network is composed of \emph{lanelets}~\cite{bender2014lanelets}, which are the fundamental units and serve as interconnected and drivable road segments. A lanelet's definition encompasses its left and right bounds, with each bound depicted by an array of points forming a polyline.
In addition, obstacles in CommonRoad contain static and dynamic obstacles, and they are characterised by their types (e.g., car, bicycle, and pedestrian) and shapes (e.g., rectangle and circle). The dynamic obstacles are also described with temporal movements in the scenario, which can be known behaviour with defined state sequence trajectory, unknown behaviour with occupancy sets, and unknown stochastic behaviour with probability distributions. 
Furthermore, each AD vehicle has a planning problem, and it is characterised by an initial state and goal region. 

Our work facilitates the automated transformation of CommonRoad scenario XML files into C-like code compatible with UPPAAL model simulations. For each scenario, we parse the XML file to extract detailed information on lanelets, obstacles, and planning problems. As shown in the green boxes in Fig.~\ref{fig:model_conversion}, we construct C-like declarations and data structures for lane boundaries, lane attributes, obstacle descriptions, planning problems, etc. By inserting these generated code blocks into appropriate sections of the template, we produce output XML files that seamlessly integrate CommonRoad scenario data into the executable UPPAAL model, thus enabling rigorous simulation and verification of driving scenarios.

\subsubsection{Drivability check: Collision and offroad detection}\label{sect:collision-offlane}
To enable a safe controller design in UPPAAL, we define the constraint on the motion planer that is guaranteed to never go off the road and collide with obstacles. Two functions, namely \texttt{offroad()} and \texttt{collide()}, are introduced to check the status of the AD vehicle and assess if the safety requirement is fulfilled.

The \texttt{offroad()} function evaluates whether a vehicle's state, represented as a rectangle or a polygon, remains within the confines of individual lanes of the lane network of a traffic scenario. Based on the geometric and kinematic relationship, the function estimates the vehicular corner points and uses them to recursively check if they are within the occupancy of a single lane, thus assessing the potential boundary violation of the lane network. 

Meanwhile, the \texttt{collide()} function checks the relationship between the AD vehicle and the other obstacles by calculating the shortest distance between two occupancy sets represented by convex polygons. The distance is calculated by circle approximation~\cite{lenz2015stochastic}, where we approximate the polygonal shape of the vehicle with a series of circles, and then compute the minimum distance between the centres of two sets of approximation circles. Given a predefined distance as the safety margin between vehicles, the minimal distance between the AD vehicle and other obstacles must fulfill this requirement, otherwise a collision happens.

\subsubsection{Behaviour models: vehicle dynamics.}
\begin{figure}[t]
    \vspace{-0.0cm}
    \setlength{\abovecaptionskip}{-0.0cm}   
    \setlength{\belowcaptionskip}{-0.5cm} 
    \centering
    \includegraphics[width=0.85\textwidth]{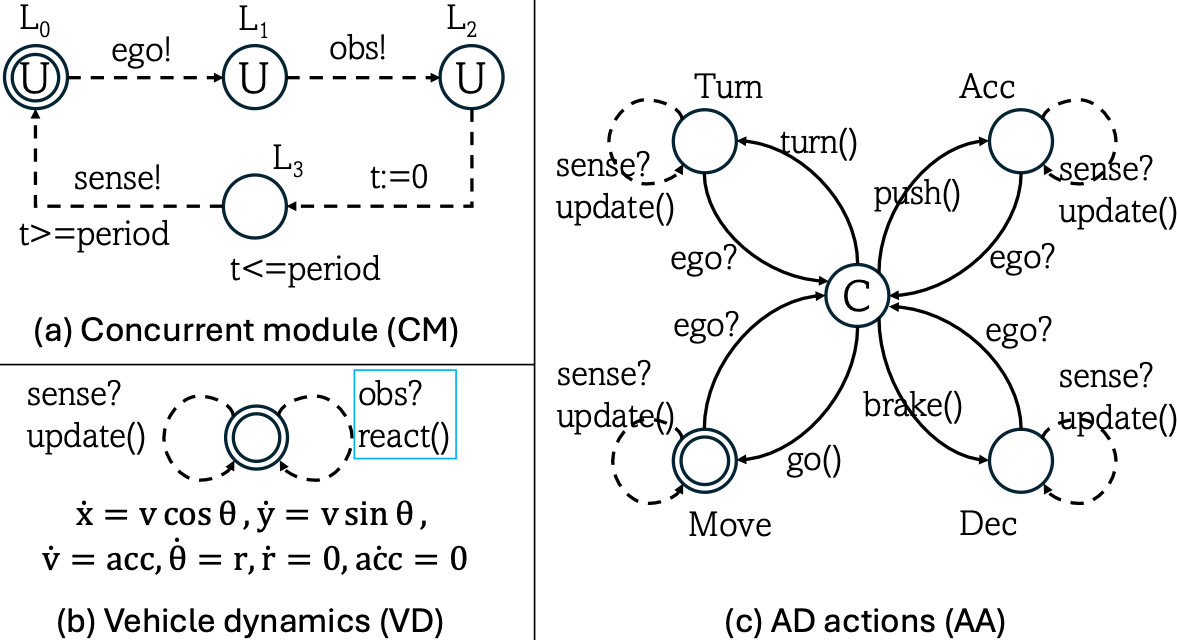}
    \caption{UPPAAL Models}
    \label{fig:uppaal_model}
\end{figure}

Vehicle dynamics and controllable actions of the AD vehicle are modelled as timed games and ordinary differential equations (ODE) in UPPAAL. Additionally, the actions of perception and controlling are synchronised. Note that the synchronisation is not mandatory, as perception and controlling as well as the actions of the AD vehicle and other vehicles can be asynchronous. Fig. \ref{fig:uppaal_model} depicts the sketch of UPPAAL models.
Model \texttt{CM} (Fig. \ref{fig:uppaal_model}(a)) is the concurrent module for action synchronisation.
It is the only timed game that deals with time, i.e., via the clock variable \texttt{t}. 
Time elapses at location \texttt{L$_3$} for exactly \texttt{period} time units, and it is followed by transitions between urgent locations, e.g., \texttt{L$_1$} and \texttt{L$_2$}, meaning that actions occur instantaneously at the end of every period.
Model \texttt{VD} (Fig. \ref{fig:uppaal_model}(b)) is a hybrid automaton describing the vehicle dynamics with ODE. The model also has two discrete transitions: one for perception, i.e., sampling the continuous variables, and one for moving obstacle reaction.
Although we include a hybrid automaton in our UPPAAL model, we can still use the symbolic model checker and search-based synthesis as long as the continuous variables are hybrid clocks, which are only used in ODE.
Note that the AD vehicle dynamics is also modelled as a hybrid automaton similar to Fig. \ref{fig:uppaal_model}(b) but without the discrete transition of moving obstacles (blue box).
Model \texttt{AA} (Fig. \ref{fig:uppaal_model}(c)) is a timed game with controllable transitions (aka, actions). It is synchronised with model \texttt{CM} via channel \texttt{ego}, meaning that the motion planner gets to choose from one of the actions in \texttt{AA} every \texttt{period} time units.
The model also has uncontrollable transitions, that is, the self-loops at locations. These transitions represent the actions of perception. Although they are performed by the AD vehicle, they do not need to be considered by the motion planner because perception happens deterministically. 
One may argue that perception may have errors that happen stochastically or arbitrarily. 
UPPAAL supports modelling such errors. However, motion planners do not need to choose among them because errors are not controllable by the AD vehicles.
In fact, the ability to model stochastic and arbitrary errors is one of the advantages of running motion planning in UPPAAL. We can even switch the semantics of errors in our UPPAAL model to enable more efficient learning than classic algorithms.
Due to the page limit, we leave this to future work.
\subsection{UPPAAL to CommonRoad}\label{sect:upp2cr}

\subsubsection{Exporting UPPAAL strategies}
A strategy in UPPAAL that its RL engine synthesises is represented as a set of decision trees over the state space. Each tree represents the Q-function for a specific action and maps a (continuous) state to the Q-value of taking the corresponding action in that state. We call this representation of a strategy a `QTree'. Such a strategy can be exported to a JSON format via the \texttt{saveStrategy} query, and we provide a Python program that can import and convert the strategy to a Python object allowing for direct interaction with CommonRoad.

Our Python program provides two classes: \texttt{QTree} and \texttt{DecisionTree}. The first is a direct adaptation of the structure of `QTree'-strategies from UPPAAL and acts as the entry point when loading a strategy. The latter is a classical (binary) decision tree with actions in its leaves. Both classes adhere to API of \texttt{stable-baselines3} \cite{stable-baselines3}, which is the default framework for most popular RL tools, such as \texttt{Gymnasium} \cite{towers_gymnasium_2023} and, relevant to this work, \texttt{commonroad-rl} \cite{itsc2021.pdf}. Most importantly, they expose a function called \texttt{predict(s)} which takes a state $s$ and returns the optimal action according to the strategy.

From a \texttt{QTree} object, one can generate a \texttt{DecisionTree} by calling the member function \texttt{to\_decision\_tree()} on the \texttt{QTree} object. The conversion happens according to a novel algorithm, that starts by sorting the leaves of all action-specific trees in the `QTree'-strategy in descending order of their Q-value. Each leaf defines a region in the state space, and since the first leaf in the aforementioned sorting has the best Q-value of all, the region defined by this leaf must be assigned the action of the tree that the leaf stemmed from. Thus, we create a shallow tree with a branch node for each upper and lower bound that makes up the leaf region. For a $k$-dimensional state space, we need $2^k$ branch nodes to perfectly capture a bounded region. Each additional leaf is added (in the sorted order) by following the path from the root node, adding new branch nodes if needed to capture the region of the leaf to be added, and discarding the addition if the leaf region is already perfectly covered by previous leaves (as these will have had a better Q-value).

The result is a decision tree that perfectly captures the strategy learned by UPPAAL\@. This allows us to return to CommonRoad after having synthesised a strategy in UPPAAL and directly interact with the tool using our safe and near-optimal control strategy. One obvious use case for this is interfacing with the RL tools provided in \texttt{commonroad-rl}, where a scenario can be executed as an RL environment. At each execution step, we obtain a state of the environment that we pass to the \texttt{predict(s)} method to get an appropriate action from the controller, which is then passed back into the environment that executes the next step. Likewise, the strategy would also be possible to use to solve motion planner problems and driveability checks.

\subsubsection{Simulating trajectories with UPPAAL}
A simpler approach than directly using the trained strategy in a CommonRoad scenario is to sample states from the strategy and show them as a trajectory in the scenario. Given the conversion from a CommonRoad scenario to a UPPAAL model, we can use the \texttt{simulate} command in UPPAAL to sample states from the execution of the model.

We store the sampled points in a log file, where concrete values of the state variables are recorded. As a default, we simply need to output the values of the five variables that determine the AD vehicle's states (coordinates on the x and y axes, orientation, velocity and acceleration) and then we add these sampled points as a trajectory of the AD vehicle to the scenario that the model was converted from. However, the user might want to add custom dynamic obstacles or special state variables in the UPPAAL model which were not part of the original scenario. Our code supports this and only requires that the extra information is added to the query file of the UPPAAL model and that the relevant constants in the Python code of model conversion are updated appropriately (e.g., update the constant of moving obstacle numbers if more are added).

We implement this process as a Python script, which provides a seamless method for visualizing the behaviour of the AD vehicle under the control of a synthesised strategy as GIF animations. 
\section{Motion Planning in UPPAAL}\label{sec:plan}
In this section, we introduce our motion-planning methods in UPPAAL. Generally, we have two classes of synthesis methods: one uses symbolic exhaustive searching and one uses RL. The former is implemented in UPPAAL Tiga \cite{behrmann2006tiga} and the latter is in UPPAAL Stratego \cite{david2015uppaal}. However, in UPPAAL 5, these two methods are integrated. 
One can synthesise a safe motion plan by using the searching method first and then optimise the safe motion plan via learning. 
The learning algorithm can be an existing function in UPPAAL, such as Q-learning \cite{watkins1992q}, or a user-defined function linked to UPPAAL as an external library \cite{gu2022correctness}.
Next, we introduce these two classes of synthesis in detail.
\subsection{Search-based Synthesis}
As the name indicates, search-based synthesis is about searching all actions at each of the reachable states and finding out the traces of state-action pairs that fulfil the requirement.
As timed games have continuous variables (i.e., clocks), their state spaces are infinite and uncountable.
Thanks to the symbolic model checker of UPPAAL, we can obtain an equivalent but discrete representation of the continuous state space, which makes the exhaustive search possible.
Since the search is exhaustive, the search-based synthesis is sound and complete, that is, when a correct motion plan exists, UPPAAL must find it (\textit{completeness}), and when UPPAAL finds a motion plan, it must be correct (\textit{soundness}).
Now that we have the model converted from CommonRoad, we can synthesise a safe motion plan that is guaranteed to never go off the road and never collide with obstacles.
The query for synthesis is as follows, where functions \texttt{collide()} and \texttt{offroad()} are defined in Section \ref{sect:collision-offlane} and provided as predefined functions in our TG templates that users of CommonUppRoad can easily reuse\footnote{More detail about the CommonUppRoad: sites.google.com/view/commonupproad}\label{ft:tool_detail}.
\begin{equation}\label{query:safe}
    \texttt{strategy safe = control: A[] !collide() $\&\&$ !offroad()}
\end{equation}
Since the search-based synthesis is complete, if UPPAAL returns \textit{unsatisfied}, it means the state space does not have any safe path. 
%
%
If Query (\ref{query:safe}) returns a strategy, it is \textit{permissive} as it contains all the safe state-action pairs, including the inefficient ones.
For example, when an AD vehicle needs to turn left at an intersection (Fig. \ref{fig:commonroad-example}), an efficient and safe strategy is to wait until the intersection is empty and then turn.
However, a permissive strategy would allow the AD vehicle to wait unnecessarily long, e.g., till the end of the motion-planning time frame, and then turn, or even go straight forward and then turn back.
Obviously, permissive strategies need to be optimised. However, they provide safety shields for RL, which gives us optimal strategies.
\subsection{Learning-based Synthesis}
Learning-based synthesis does not need to exhaustively search the entire state space. Instead, it randomly simulates the model and samples traces of state-action pairs.
Note that the sampled state-action pairs are not symbolic anymore, as the state-space exploration now is a random simulation.
Therefore, learning-based synthesis is not sound or complete. We need methods to provide safety guarantees on the learning results.
Previously, we have proposed a post-verification on the learning results \cite{gu2022correctness}.
One can use this method by running the following queries, where \texttt{MAXT} is the maximum time of one learning episode, \texttt{reward} is the formula representing the \textit{reward function} of RL, and \texttt{goal()} is a function returning \textit{true} when the AD vehicle reaches the goal.
\begin{equation}\label{query:reach}
    \texttt{strategy reach = maxE(reward)[<=MAXT]: <> goal()}
\end{equation}
\begin{equation}\label{query:verireachSafe}
    \texttt{A[] !collide() $\&\&$ !offroad() under reach}
\end{equation}
\begin{equation}\label{query:verireachGoal}
    \texttt{A<> goal() under reach}
\end{equation}
Query (\ref{query:reach}) returns a strategy that \textit{may} reach the destination.
The reachability and safety of strategy \texttt{reach} are not guaranteed as learning is based on random simulation.
Hence, we use Queries (\ref{query:verireachSafe}) and (\ref{query:verireachGoal}) to verify the strategy \texttt{reach}.
The former checks if the AD vehicle \textit{never} collides and \textit{never} goes off the road under the control of strategy \texttt{reach}. The latter checks if the AD vehicle \textit{always eventually} reaches the goal regardless of other vehicles' actions.
Learning-based synthesis can be combined with search-based synthesis by using the following query, where \texttt{safe} is a strategy synthesised by Query (\ref{query:safe}). Hence, Query (\ref{query:safe}) must be executed prior to running this query.
\begin{equation}\label{query:reachSafe}
    \texttt{strategy reachS = maxE(reward)[<=MAXT]: <> goal() under safe}
\end{equation}
The random simulation of Query (\ref{query:reachSafe}) is restricted to the state-action pairs included in the permissive strategy \texttt{safe}.
Hence, the simulation would never reach \textit{unsafe} states and thus Query (\ref{query:verireachSafe}) is not needed.
Strategy \texttt{safe} performs as a safety shield of the learning process. 
Similar to Query (\ref{query:reach}), although Query (\ref{query:reachSafe}) explicitly specifies the learning objective is to reach the goal states, \texttt{reachS} is not guaranteed to be goal-reaching, because learning has no correctness guarantee.
For example, the learning episodes can be not enough to cover all the branches that the state space has, or the reward function is problematic.
%
In addition, the time for running Query (\ref{query:safe}), the precondition of running Query (\ref{query:reachSafe}), can be extremely long as it requires an exhaustive state space exploration. 
However, when the AD actions for the motion planner to choose are not too many but the distance of travelling is long, learning-based methods may perform worse than the search-based methods.
We have evaluated and reported the strengths and weaknesses of both methods in multi-agent planning \cite{gu2022verifiable}.
\section{Experiments}\label{sec:exp}
We have conducted two groups of experiments to show the usability and scalability of CommonUppRoad, respectively. We run the experiments on an M2-chip Macbook Pro with 16 GB memory.
In the following sections, we introduce the design of the experiments and the results$^3$. 
\subsection{Experiment Design}
%
\begin{figure}[t]
    \vspace{-0.0cm}
    \setlength{\abovecaptionskip}{-0.0cm}   
    \setlength{\belowcaptionskip}{-0.5cm} 
    \centering
    \includegraphics[width=0.9\textwidth]{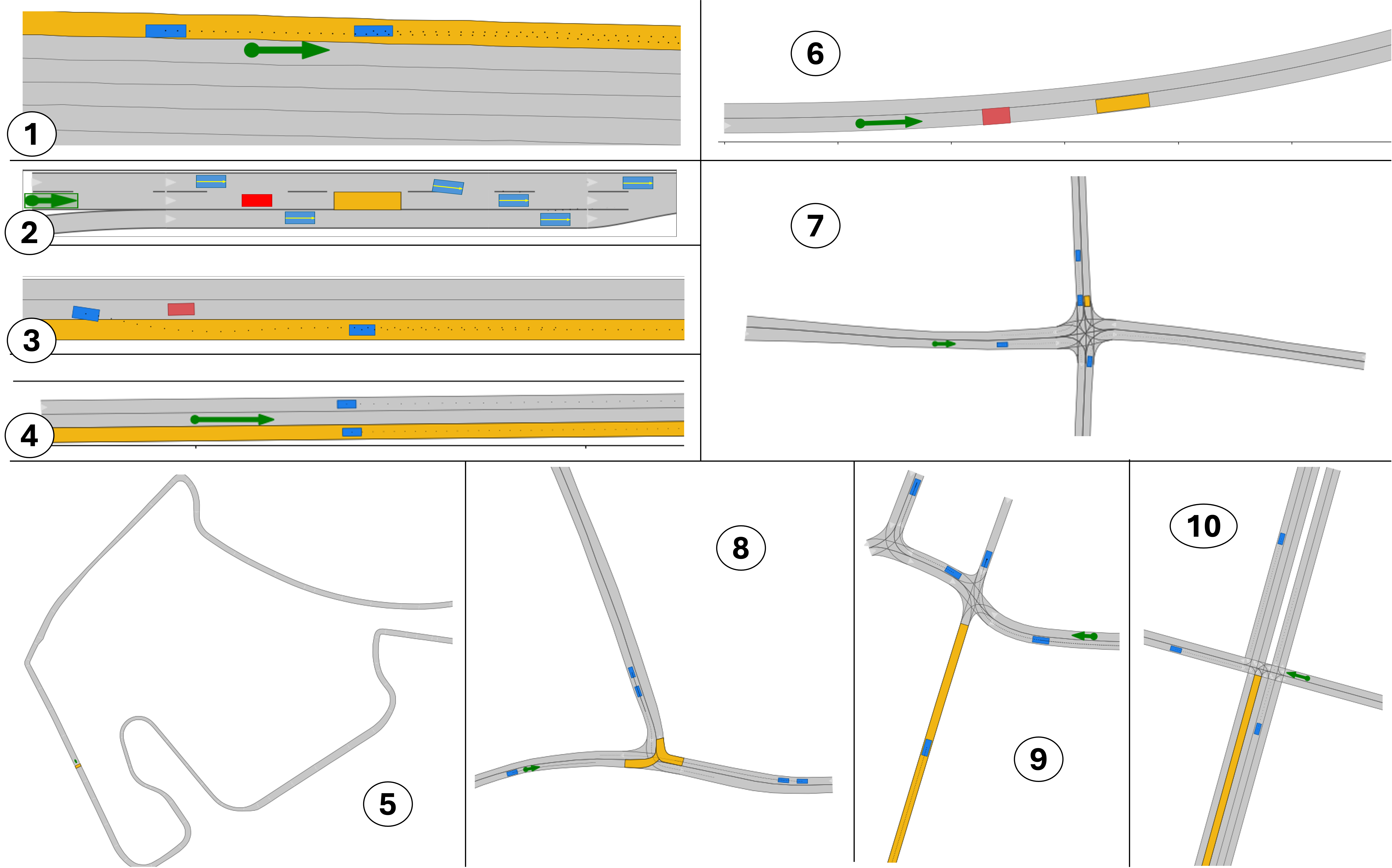}
    \caption{Scenarios for the experiments.}
    \label{fig:exp1_scenarios}
\end{figure}
\textbf{Experiment I} (usability): We select ten representative scenarios in CommonRoad (Fig. \ref{fig:exp1_scenarios}), from simple ones to complex ones, and convert them to UPPAAL models by using a Python program implementing the method in Section \ref{sect:cr2upp}.
To tame the sizes of the models, we define the maximum time to be ten, meaning that the models are forced to reset to the initial state when the execution time turns ten.
This restriction of execution time would limit the model sizes within a reasonable scale.
We further evaluate the scalability in Experiment II, where the model sizes can be much larger.
We check the syntax of the generated UPPAAL models, proving the syntactic correctness of the generated models. Then, we verify the following query in UPPAAL. If the query returns \textit{true}, the model exists a trace where the AD vehicle never collides or goes off the road.
\begin{equation}\label{query:exist}
    \texttt{E[] !collide() $\&\&$ !offroad()}
\end{equation}
Last, we run Queries (\ref{query:safe}) and (\ref{query:reachSafe}) to generate safe motion plans. Note that since the execution time is limited, the starting points of the AD vehicles are set to be close to the goals such that they are reachable within the time limit.
However, Experiment I is about showing the usability of CommonUppRoad, which is supported by the experiment result already.
\textbf{Experiment II} (scalability): We select scenario \circled{2} in Fig. \ref{fig:exp1_scenarios} and modify it in this experiment.
We change the maximum time steps for a learning episode and the distance from the starting point to the goal area accordingly.
We run Queries (\ref{query:safe}) and (\ref{query:reachSafe}) to show how scalable the search-based method is. Then, we run Query (\ref{query:reach}) and verify the resulting strategies against Queries (\ref{query:verireachSafe}) and (\ref{query:verireachGoal}) to show the correctness of the learning-based method. 
\subsection{Experiment Results}
Table \ref{tab:res-exp1} shows the result of Experiment I. All the converted UPPAAL models pass the syntax check, which shows the static information of scenarios can be successfully transferred to variable declarations in UPPAAL.
Further, the results of queries are all \textit{satisfied} (\checkmark) meaning that the UPPAAL model can generate safe and reachable motion plans. 
In general, the computation times of all scenarios are reasonable (less than 30 seconds) because we set the goals to be close to the starting points.
Note that the two moving obstacles are near the AD vehicle in Scenario \circled{1} but are far from the AD vehicle in other scenarios when the latter approaches the goal.
Therefore, the computation of the Query (\ref{query:safe}) is the longest in Scenario \circled{1}, which also indicates the largest safe strategy.
Consequently, the computation time of the other two queries is also the longest in Scenario \circled{1}.
The result of Experiment I indicates the possibility of compositional motion planning in complex scenarios where scalability becomes an issue.
\begin{table}[t]
\centering
\caption{Results of Experiment I}
\label{tab:res-exp1}
\resizebox{0.65\textwidth}{!}{%
\begin{tabular}{|c|c|cc|cc|cc|}
\hline
\multirow{2}{*}{Scenario} & \multirow{2}{*}{Syntax} & \multicolumn{2}{c|}{Query (\ref{query:exist})}  & \multicolumn{2}{c|}{Query (\ref{query:safe})}  & \multicolumn{2}{c|}{Query (\ref{query:reachSafe})}  \\ \cline{3-8}
                &            & \multicolumn{1}{c|}{SAT} & Time (ms) & \multicolumn{1}{c|}{SAT} & Time (ms) & \multicolumn{1}{c|}{SAT} & Time (ms) \\ \hline
\circled{1}     & \checkmark & \multicolumn{1}{c|}{\checkmark} & \textbf{3,186} & \multicolumn{1}{c|}{\checkmark} & \textbf{28,928} & \multicolumn{1}{c|}{\checkmark} & \textbf{24,628}     \\ \hline
\circled{2}     & \checkmark & \multicolumn{1}{c|}{\checkmark} & 65 & \multicolumn{1}{c|}{\checkmark} & 2,264 & \multicolumn{1}{c|}{\checkmark} & 5,090     \\ \hline
\circled{3}     & \checkmark & \multicolumn{1}{c|}{\checkmark} & 4 & \multicolumn{1}{c|}{\checkmark} & 315 & \multicolumn{1}{c|}{\checkmark} & 1,337     \\ \hline
\circled{4}     & \checkmark & \multicolumn{1}{c|}{\checkmark} & 111 & \multicolumn{1}{c|}{\checkmark} & 3,473 & \multicolumn{1}{c|}{\checkmark} & 8,132     \\ \hline
\circled{5}     & \checkmark & \multicolumn{1}{c|}{\checkmark} & 333 & \multicolumn{1}{c|}{\checkmark} & 142 & \multicolumn{1}{c|}{\checkmark} & 13,842     \\ \hline
\circled{6}     & \checkmark & \multicolumn{1}{c|}{\checkmark} & 335 & \multicolumn{1}{c|}{\checkmark} & 14,364 & \multicolumn{1}{c|}{\checkmark} & 21,796     \\ \hline
\circled{7}     & \checkmark & \multicolumn{1}{c|}{\checkmark} & 179 & \multicolumn{1}{c|}{\checkmark} & 12,024 & \multicolumn{1}{c|}{\checkmark} & 12,667     \\ \hline
\circled{8}     & \checkmark & \multicolumn{1}{c|}{\checkmark} & 127 & \multicolumn{1}{c|}{\checkmark} & 55 & \multicolumn{1}{c|}{\checkmark} & 6,523     \\ \hline
\circled{9}     & \checkmark & \multicolumn{1}{c|}{\checkmark} & 124 & \multicolumn{1}{c|}{\checkmark} & 54 & \multicolumn{1}{c|}{\checkmark} & 6,444     \\ \hline
\circled{10}    & \checkmark & \multicolumn{1}{c|}{\checkmark} & 359 & \multicolumn{1}{c|}{\checkmark} & 154 & \multicolumn{1}{c|}{\checkmark} & 15,362     \\ \hline
\end{tabular}%
}
\end{table}
\begin{wrapfigure}[10]{l}{0.38\columnwidth}
\vspace{-1.2cm}
\setlength{\abovecaptionskip}{-0.00cm}   
\setlength{\belowcaptionskip}{-0.00cm} 
\begin{center}
  \includegraphics[width=0.38\columnwidth]{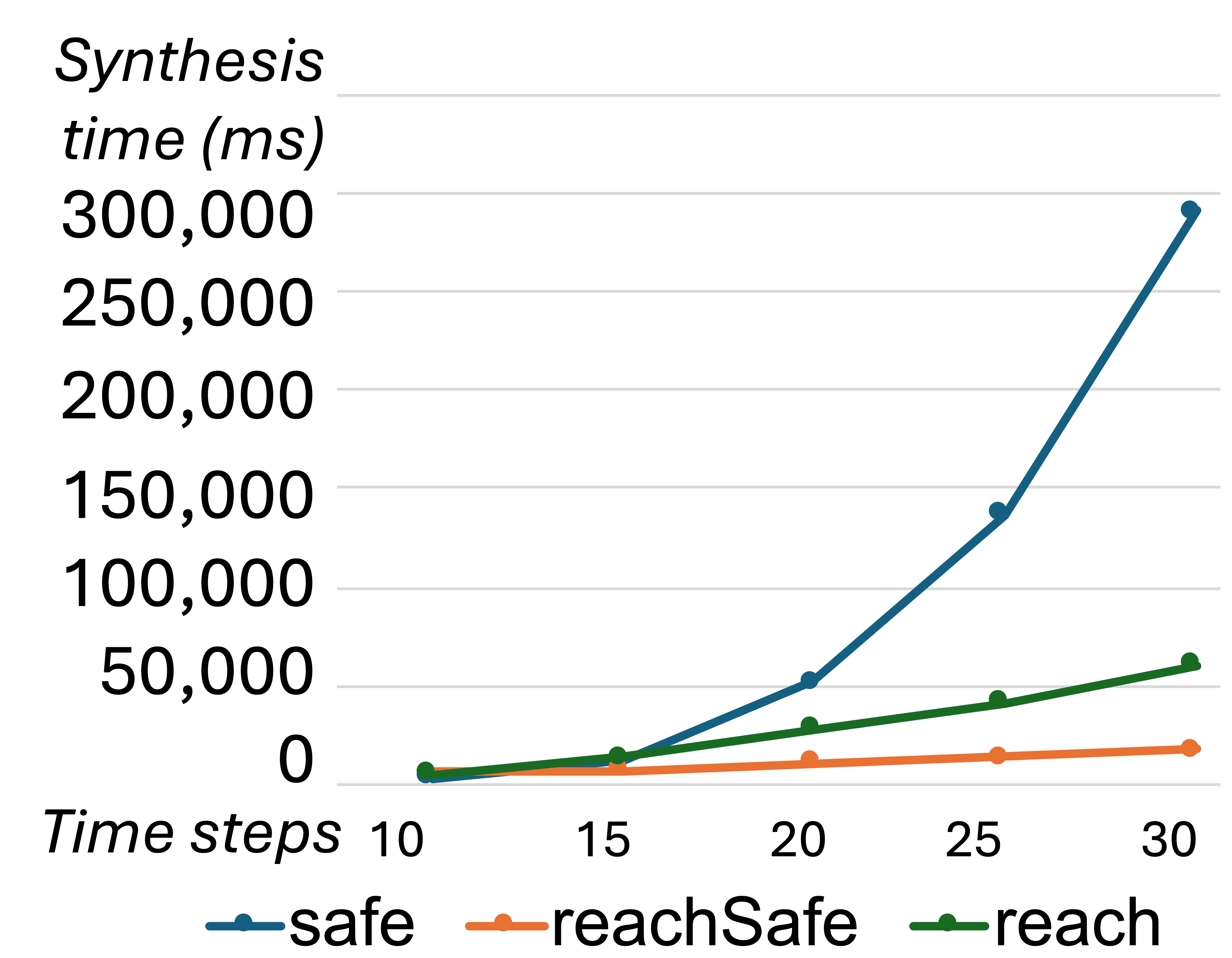}
  \caption{Experiment II Result}
  \label{fig:exp2_result}
\end{center}
\end{wrapfigure}
Figure \ref{fig:exp2_result} shows the trends of computation times for different kinds of motion plans.
As expected, the time for computing safe motion plans (i.e., \texttt{safe}) increases exponentially as they are synthesised by the search-based method.
However, the time for optimising the safe plans (i.e., \texttt{reachSafe}), increases linearly as the synthesis method is learning and the state space is restricted by the safe plans already.
The motion plans that are synthesised purely by learning (i.e., \texttt{reach}) do not have safety and reachability guarantees.
Therefore, we verify the learning results against Queries (\ref{query:verireachSafe}) and (\ref{query:verireachGoal}).
When the verification returns \textit{false},  we increase the learning episodes and learn again until the results satisfy those queries.
Figure \ref{fig:exp2_result} shows the computation time for the final round of learning whose verification results are \textit{true}.
Although the computation times of the learning-based method are shorter than those of the search-based method, it involves a trial-and-error process, that is, repetitive learning when verification fails. 
If we count this time in, the entire synthesis time can be even longer than the time of the search-based method.
However, when the model state space becomes too big to be handled by the search-based method, learning is a good solution.
The animated visualisation of UPPAAL-synthesised motion plans is posted online$^3$.
\subsection{Discussion}
We answer the research questions in Section \ref{sec:problem} based on the experiment results.
\begin{tcolorbox}[width=\textwidth,colback={anti-flashwhite},title={},colbacktitle=antiquewhite,coltitle=blue]    
Answer to \textbf{RQ1}: Scenarios in CommonRoad can be converted to constant variables defined in the timed-game templates in UPPAAL. CommonRoad functions of collision-detection and offroad-detection can be converted to the C-like functions in UPPAAL.
\end{tcolorbox} 
Although the correctness of model conversion has been shown in the experiments, there is room for improvement. For example, offroad detection in UPPAAL is done over the entire road network, which is not always necessary and influences efficiency. 
%
\begin{tcolorbox}[width=\textwidth,colback={anti-flashwhite},title={},colbacktitle=antiquewhite,coltitle=blue]    
Answer to \textbf{RQ2}: UPPAAL strategies can be converted to decision-tree controllers in CommonRoad, or sampled points of trajectories that are visualised in CommonRoad.
\end{tcolorbox} 
In the experiments, we have demonstrated the visualisation of possible trajectories that a decision-tree controller can have. However, there is no completeness guarantee on the sampled trajectory points, because sampling is via random simulations.
As our timed-game templates preserve the ability of exhaustive verification, one way of collecting all the possible trajectories of a decision-tree controller is to use \textit{liveness} properties in UPPAAL. Specifically, during the verification of Query (\ref{query:verireachGoal}), we can label the state-action pairs visited by the model checker and sample concrete states in the labelled symbolic states.
This process is similar to our strategy compression proposed previously \cite{gu2022correctness}. However, due to the page limit, we leave this as future work.
\begin{tcolorbox}[width=\textwidth,colback={anti-flashwhite},title={},colbacktitle=antiquewhite,coltitle=blue]    
Answer to \textbf{RQ3}: The search-based method can generate safe motion plans and be used as safety shields for learning. However, its computation time increases exponentially as the time step of the AD vehicle increases linearly. In contrast, the computation time of the learning-based method increases linearly. However, the learning results do not have a safety guarantee. Post-verification of the learning results would overcome this limit, but it involves a trial-and-error process, which can be time-consuming.
\end{tcolorbox} 
As we reported previously \cite{gu2022verifiable}, one shall adaptively employ the search-based and learning-based methods according to their applications. Compositional motion planning can be a good solution for the scalability issue. However, it involves extra effort such as contract-based concatenation. We leave this for future work.
\section{Related Work}\label{sec:related}
To ensure the reliability and safety of autonomous driving (AD) systems, formal methods (FMs) are widely applied for the verification and validation of motion planning algorithms. Researchers use formal specifications to define safety properties and other constraints that the planning algorithms must satisfy~\cite{mehdipour2023formal}.

Specifically, CommonRoad has been used as a platform for applying FMs, such as temporal logic, to solve various aspects of motion-planning problems.
Lercher et al.~\cite{lercher2024specification} use linear temporal logic specification for the reachability analysis to overapproximate the reachable set of an AD vehicle and validate the implementation in the CommonRoad platform.
%
%
Liu et al.~\cite{liu2023specification} address specification-compliant motion planning for AD vehicles based on set-based reachability analysis with automata-based model checking. The effectiveness of the methods is demonstrated with scenarios from the CommonRoad benchmark suite.
Hekmatnejad et al.~\cite{hekmatnejad2019encoding} translate Responsibility-Sensitive-Safety (RSS) rules into Signal Temporal Logic (STL) formulas and utilise the STL formulas to monitor off-line naturalistic driving data provided with CommonRoad.
In comparison, our framework provides a bi-directional model conversion between CommonRoad and UPPAAL.
In addition, CommonUppRoad has two ways of motion planning, i.e., search-based and learning-based, and the corresponding verification methods.
These features make our framework comprehensive and greatly benefit control engineers unfamiliar with FMs.
%
%
%
%

%
In the FMs community, researchers have studied ways to facilitate system engineers to apply FMs in AD systems.
Gu et al. \cite{gu2022malta} develop a tool-supported methodology for AD vehicle mission planning, namely MALTA, in which UPPAAL is employed as a task scheduler at the backend of the tool.
%
%
Besides UPPAAL, there are many other tools that can potentially solve the AD motion-planning problem, e.g., Kronos \cite{bozga1998kronos}, LTSim \cite{blom2010ltsmin}, and SpaceEx \cite{frehse2011spaceex}.
These tools mainly suffer from a common problem: state-space explosion. Compositional planning has the potential to overcome the state-space explosion as it can split the entire problem into smaller but compositional sub-problems \cite{alur2018compositional}\cite{muhammad2024energy}.
%
%
%
Our experiment shows the ability of CommonUppRoad to be used in compositional planning. The well-defined structure of scenarios provides the foundation for composing contracts for concatenating motion plans of sub-problems.
\section{Conclusion and Future Work}\label{sec:conclusion}
We propose a framework called \textit{CommonUppRoad} that combines CommonRoad and UPPAAL. The framework provides automatic model conversion between CommonRoad and UPPAAL, which would greatly benefit control engineers to adopt formal methods in their system design, analysis, and verification.
We also offer timed-game (TG) templates to describe the vehicle dynamics and C-like functions of collision- and offroad-detection. These features would facilitate learning algorithm designers to use formal methods, that is, TG as the modelling language, and (statistical) model checking as the verification technique, in the domain of AD vehicles.
We report the performance of search-based and learning-based synthesis in various scenarios. The search-based method can generate safe motion plans but scales badly when the model becomes large and complex. The learning-based method scales better, but it does not have a safety guarantee. Post-verification can compensate for this shortcoming, but it involves a trial-and-error process to obtain the final safety-guaranteed result.
One of the future works is to investigate compositional motion planning, which would overcome the state-space explosion profoundly. 
Another direction is to use model checking in UPPAAL to interpret AD scenarios (aka, operational design domain (ODD)) and generate critical scenarios that are highly likely to cause collisions. This scenario generation and selection method would greatly benefit the industry in AD testing, verification, and validation.
\bibliographystyle{splncs04}
%

%
\end{document}